# Persistent Gaps, Partial Gains: A Population-Level Study of COVID-19 Learning Inequalities in the Netherlands

**Hekmat Alrouh, Tom Emery, and Anja Schreijer**

__**Abstract**__

The COVID-19 pandemic disrupted schooling worldwide, raising concerns about widening educational inequalities. Using population-level administrative data from the Netherlands (N = 1,471,217), this study examines how socio-economic disparities in secondary school performance evolved before, during, and after pandemic-related school closures. We analyze final central examination scores for cohorts graduating between 2017 and 2023 across four educational tracks, estimating generalized linear models with interactions between pandemic exposure and key stratification variables: parental education, household income, migration background, and urbanicity. Results show that while average performance partially recovered by 2023, inequalities by parental education and migration background persisted or intensified, particularly in vocational tracks. First-generation students with a non-Western background experienced the largest sustained losses, whereas students in rural areas—previously disadvantaged—narrowed or reversed pre-pandemic performance gaps. Findings suggest that systemic shocks can both exacerbate and recalibrate inequality patterns, depending on the socio-demographic dimension and educational context. We discuss implications for stratification theory, highlighting the role of educational pathways and local contexts in shaping resilience to crisis-induced learning disruptions.

# 1. INTRODUCTION

## COVID-19 School Closures and Learning Losses

The COVID-19 pandemic precipitated an unprecedented disruption of education worldwide. At the peak of the crisis in 2020, an estimated 95% of the global student population was affected by school closures or shifts to remote learning (United Nations, 2020). A growing body of research has since documented significant learning losses in the form of reductions in academic progress or achievement attributable to these disruptions (Betthäuser et al., 2023; Hammerstein et al., 2021). Early studies reported measurable declines in student performance even in contexts with relatively short lockdowns and robust infrastructure. For instance, even in the pandemic's first year, students made little or no progress with an average learning deficit of ~0.08 standard deviations, roughly equivalent to one-fifth of a school year of learning (Engzell et al., 2021). This finding was echoed in results from other countries and prompted widespread concern that the pandemic had set back an entire cohort of learners. Indeed, a recent meta-analysis of 42 studies across 15 countries estimates an overall learning shortfall of about 0.14 standard deviations, which arose early in the pandemic and persisted over time (Betthäuser et al., 2023). Using PISA trend data through 2022, Jakubowski *et al.* estimate an average decline in mathematics of ~0.14 SD—roughly seven months of learning—with steeper losses in systems that experienced longer school closures (Jakubowski et al., 2025).

In early-tracking systems like the Netherlands, pandemic disruptions also touched high-stakes transitions. Dutch register data show that while tracking outcomes (teacher recommendation and placement in grades 7 and 9) were less affected than test scores in other studies, adverse effects concentrated among students from disadvantaged backgrounds. Importantly, students' academic motivation, self-efficacy, and parental involvement were associated with smaller negative pandemic effects on these transitions, pointing to the buffering role of academic and social embeddedness when schooling is disrupted (van de Werfhorst et al., 2024).

A recent systematic review covering 57 studies across 38 countries provides further nuance by disentangling the effects of school closure from the effects of reopening (Dela Cruz et al., 2025). The review estimates that one year of school closure corresponded to learning losses equivalent to 1.1 years of schooling, while reopening schools reduced this to about 0.5 years. This underscores both the severity of the initial disruption and the mitigating effect of restoring in-person instruction. Importantly, the review shows that recovery after reopening has been only partial, with learning deficits persisting well into the post-pandemic period. Losses were particularly large in mathematics and science compared to reading, and greater for primary school students than for those in secondary education. Moreover, although schools in developing countries were closed for nearly twice as long as those in high-income countries, the observed gap in learning loss was smaller than expected, reflecting both differences in baseline education quality and the uneven mitigating effects of reopening. These findings highlight that reopening alone cannot erase losses; targeted and sustained recovery measures are required to prevent long-term scarring of cohorts affected by the pandemic

**Widening Socio-Economic Gaps**

Crucially, these learning losses have not been distributed evenly; instead, COVID-19 appears to have exacerbated pre-existing educational inequalities (Engzell et al., 2021; Maldonado & De Witte, 2022). Evidence from multiple studies indicates that students from socio-economically disadvantaged backgrounds were hit the hardest. In the Netherlands, learning losses were up to 60% larger for children from less-educated households (Engzell et al., 2021). Another large study of Dutch pupils found disproportionately greater learning deficits among students from lower parental education and income groups, on top of the inequalities that already existed pre-pandemic (Haelermans, Korthals, et al., 2022).

Evidence from the United States highlights how the pandemic amplified pre-existing disparities. Large multi-district datasets show that Black, Hispanic, and American Indian/Alaska Native students, and students in high-poverty schools, experienced disproportionately larger drops in reading and mathematics during 2020–21, with gaps widening relative to White and Asian peers (Gee et al., 2023). As in-person schooling resumed, achievement trajectories rebounded system-wide but remained notably lower in high-poverty schools, indicating incomplete recovery among the most disadvantaged. The pandemic period also saw sharp increases in absenteeism that disproportionately affected Black and Hispanic students, reflecting both health-related quarantines and structural barriers such as housing instability and limited transportation (Gee et al., 2023).

Results from South Africa's nationally representative PIRLS provide an LMIC counterpoint: Grade 4 reading declined by 31 points (≈0.29 SD; ~50–60% of a school year), with widening dispersion both nationally and within schools (Böhmer & Wills, 2025). The share of very low achievers (below 200 PIRLS points) doubled, and a notable rise in non-attempted items suggests disengagement and severe instructional discontinuities. Losses were concentrated among students in the poorest 70% of schools and were larger for boys, illustrating how prolonged rotation/partial attendance models can entrench within-school inequality when instructional time is unevenly restored (Böhmer & Wills, 2025).

These patterns align with international reports: for example, a cross-country meta-analysis concludes that learning deficits were particularly large among children from low socio-economic backgrounds (Betthäuser et al., 2023). Results from PISA also showed larger losses for boys, immigrant students, and socio-economically disadvantaged youth (Jakubowski et al., 2025).

As the pandemic evolved, attention turned to whether students could recover lost learning and how to support that recovery. By late 2020 and 2021, governments worldwide launched extensive learning recovery programs. In a recent study, secondary schools, students who participated in tutoring or remedial programs showed significantly greater improvements in their grades compared to similar peers who did not participate, indicating that intensive support helped to partially offset the learning losses from school closures (de Bruijn & Meeter, 2023).

Despite these efforts, full recovery has been elusive at the system level. One year or more after the initial lockdowns, students on average had *not* fully regained the ground lost to the pandemic (Di Pietro, 2023). Meta-analyses conclude that learning deficits persisted into the 2021–2022 school year, with losses in subjects like mathematics remaining significant and no clear evidence of self-correction of learning trajectories (Di Pietro, 2023). In other words, without substantial interventions, the passage of time alone has not closed the gaps created during the pandemic. This raises pressing concerns about the long-term implications of COVID-related learning delays. Educational attainment is cumulative; missing foundational skills in secondary school can hinder students' progress to higher education and limit their skill development for the labor market. Researchers warn that, if unresolved, the COVID-19 learning crisis could cast a long shadow on this generation's prospects (Betthäuser et al., 2023; Psacharopoulos et al., 2021). Modelled projections suggest that lower learning levels may translate into reduced lifetime earnings and productivity, especially for those from poorer backgrounds, who have less familial safety net to compensate (Maldonado et al., 2024; World Bank Group, 2021). There are also worries about social consequences: widened educational disparities can undermine social cohesion and economic equity in the long run (Maldonado et al., 2024). These considerations underscore that learning loss is not merely a short-term scholastic issue, but a matter of educational justice and future opportunities for disadvantaged youth.

**The impact of COVID on schools in the Netherlands**

The Netherlands implemented three nationwide school closures affecting secondary education (voortgezet onderwijs) during the COVID-19 pandemic. The first closure ran from March 16 to June 2, 2020, during which all secondary teaching shifted fully online. A second closure lasted from December 16, 2020 until February 8, 2021, followed by a partial reopening. The third closure (essentially an extended winter break) occurred from December 14, 2021 to January 10, 2022. Outside these periods, schools generally reopened under strict conditions. Even when schools were open, many students and staff faced intermittent quarantine disruptions, causing ongoing learning interruptions. Notably, the 2020 nationwide central exams for secondary school leavers were canceled; students graduated based on school assessments instead. In subsequent years, exam schedules and grading were adjusted to accommodate pandemic disruptions.

Once schools reopened, comprehensive mitigation measures were in place per government and RIVM guidelines, including an initial 1.5-meter distancing rule applied in secondary schools. To comply, schools split classes and rotated attendance (e.g. half the students attending on different days). This hybrid model, used in mid-2020 and again in early 2021, reduced contacts while delivering part-time in-person instruction. Even after full classes resumed in June 2021, students no longer had to distance from each other, though they were advised to keep distance from teachers.

Face masks were introduced as a requirement in autumn 2020. From October 2020, all secondary students and staff had to wear masks outside of classrooms (e.g. in hallways). During later waves (2021), mask rules intensified – for example in late November 2021 masks became mandatory in corridors for all, alongside regular self-testing. Mask mandates in schools remained in effect through early 2022 and were lifted once distancing rules ended in February 2022. Starting in 2021, the government supplied rapid antigen self-test kits to secondary schools. By spring 2021, students and teachers were urged to self-test twice per week at home, especially once full attendance resumed. In late 2021 this became a strong recommendation for all secondary students and staff. Teachers were given priority access to PCR testing from September 2020 to minimize sick-leave gaps. Strict protocols required anyone with respiratory symptoms to stay home and get tested, with schools providing remote learning for those in isolation.

COVID-19 vaccinations for adolescents began in mid-2021. From early July 2021, all youth aged 12–17 were eligible for the Pfizer vaccine, with the aim that willing students received a first dose before the new school year. Vaccination was encouraged (informational campaigns were provided in schools) but not mandated for attendance. Teachers were not prioritized by age in the vaccine rollout (the cabinet opted not to vaccinate educators ahead of their age group), though most had the opportunity to be vaccinated by summer 2021.

COVID-19 measures significantly affected educational outcomes in Dutch secondary education. Recurring periods of remote learning and hybrid instruction led to concerns about learning loss and student well-being. Studies in 2020 suggested that even an 8-week closure resulted in measurable learning delays, with disadvantaged students falling furthest behind. By mid-2021 (after ~16 months of disruptions), school leaders reported greater concern for students' motivation, mental health, and socio-emotional development than for academic gaps (Ministry of Education, Culture and Science, 2021). Older pupils – who endured longer stretches of online learning – and those from unstable home environments were most adversely impacted. A majority of surveyed secondary students admitted to being less motivated and putting in less effort during the pandemic; nearly half worried about learning backlogs, and a quarter feared they might never catch up (Ministry of Education, Culture and Science, 2021). In response, the Dutch government and educational authorities took steps to mitigate learning deficits. A National Program for Education (Nationaal Programma Onderwijs) was launched in early 2021, investing €8.5 billion over 2.5 years in tutoring, remedial teaching, summer schools and mental health support to help students recover academically and emotionally. Secondary schools used these funds to provide additional lessons and support aimed at reversing learning loss. Crucially, the experience of 2020–2021 influenced policy: by 2022 the government and the VO-raad (secondary education council) agreed that even in worst-case pandemic scenarios, schools should avoid full closure. Instead, “red” scenario plans would keep students attending at least 50% of the time in person (through shifts or hybrid models), reflecting a commitment to maintain continuity of education. This emphasis on keeping secondary schools open, alongside targeted support programs, aimed to safeguard students’ educational progress and well-being in the face of COVID-19.

**Theories of Inequalities in learning loss and recovery**

The recovery of educational outcomes following pandemic-related school closures has been uneven, with differences in both rate and extent of rebound across social groups. A range of theoretical perspectives can be used to generate specific hypotheses about the trajectory of recovery curves over the post-pandemic period, with implications for whether inequalities narrow, persist, or widen.

Education production and time-on-task theories (Hanushek, 1979) emphasise the role of instructional time and resource inputs in generating learning. From this perspective, we expect that most students will show year-on-year gains after schools reopen, gradually approaching pre-pandemic trajectories. However, recovery will be steeper for students with access to more effective teachers, tutoring, and supportive learning environments, producing convex recovery curves for advantaged students and flatter curves for disadvantaged peers. Without resource equalisation, inequalities are likely to persist. Parents in upper-middle class households often had the flexibility and resources to supervise their children's remote schooling, whereas working-class parents (many serving as essential workers) could not stay home to assist with schoolwork (Chen et al., 2022). These factors likely also apply post-pandemic.

Cumulative (dis)advantage and Matthew effects (DiPrete & Eirich, 2006; Stanovich, 1986) highlight the path-dependent nature of skill acquisition. Students with stronger pre-pandemic achievement will recover more quickly, because foundational skills enable faster uptake of new material. Lower-achieving students will recover slowly and may plateau, leading to diverging trajectories: advantaged groups show rapid initial recovery while disadvantaged groups' gains are limited. As cohorts who spent their formative years of schooling pass through the graduation phase of their education, we should observe persistent or even widening gaps in educational outcomes when compared to pre-pandemic and particularly amongst students from a lower socio-economic background.

From a cultural capital and inequality reproduction perspective (Bourdieu, 1983; Lareau, 2003), high-SES parents can better navigate recovery resources—securing tutoring, advocating for accommodations, and maintaining structured learning routines. Both high- and low-SES students improve over time, but the vertical gap between them remains stable or even slightly larger by year producing parallel upward-trending curves. So whilst recovery is universal, the inequality is likely persistent.

Relative risk aversion, Effectively Maintained Inequality (Lucas, 2001), and compensatory advantage (Bernardi & Triventi, 2020) suggest that advantaged families will act quickly to restore their children's pre-pandemic standing, making sharp early gains that protect high-status credentials. Once pre-pandemic gaps are re-established, recovery in these groups plateaus, while disadvantaged students' curves rise more gradually through recovery.

A social capital and ecological systems lens (Bronfenbrenner, 1979; Coleman, 1988) shifts the focus from individual to community resources. Schools embedded in dense networks of trust and cooperation—whether advantaged or disadvantaged—can accelerate recovery through communication, coordination, and targeted support. In such contexts, disadvantaged groups may show concave recovery curves, with slower initial gains but acceleration over time, potentially narrowing gaps over time as resources are mobilized to compensate for material disadvantages.

The digital inequality framework (Hargittai, 2002; van Dijk, 2005) predicts a delayed recovery for students who lacked digital access, skills, or support during closures. The abrupt shift to online instruction amplified the digital divide, as disadvantaged students were less likely to have access to adequate devices, high-speed internet, and a quiet place to study (Golden et al., 2023). Consequently, students from less advantaged families not only had fewer technological resources for online learning, but also received less direct support and oversight during the lockdown (Golden et al., 2023). These disparities led to markedly different learning experiences: one study noted that during the first lockdown, more vulnerable student populations saw significantly larger drops in academic progress than those with more affluent backgrounds (Haelermans, Jacobs, et al., 2022). These students start further behind and recover slowly but the gap narrows as in-person learning resumes dominance, but a residual gap remains due to missed opportunities to build digital competence.

An intersectional perspective (Crenshaw, 1989) posits that recovery will be slowest and flattest where multiple disadvantages intersect—such as low-SES, migrant-background girls in vocational tracks. These students face compounding barriers, resulting in persistently low recovery slopes and widening disparities relative to all other groups.

Tracking and school stratification theories argue that system-level sorting into high- and low-status educational tracks shapes recovery and there is considerable evidence of this being increasingly along socio-economic lines during the pandemic (van de Werfhorst et al., 2024). Higher-status tracks recover rapidly within the first one to two years due to stronger curricula, more experienced teachers, and selective intakes, while lower-status tracks recover slowly and incompletely, producing persistent vertical separation in recovery curves.

From a capability approach perspective (Nussbaum, 2011; Sen, 1999), what matters is not just resources but the real freedoms learners have to convert them into learning. Students facing mental health challenges, unsafe home environments, or caregiving burdens may show erratic or wavy recovery curves, with potential setbacks interrupting gains, even when attending reopened schools.

Finally, theories of policy design and targeting stress that the shape of recovery is contingent on intervention. In systems implementing targeted, high-dosage supports (e.g., tutoring for the most affected), disadvantaged groups' recovery curves may steepen, leading to eventual convergence and full recovery. Without targeting, advantaged groups maintain or expand their lead, and curves remain parallel.

Taken together, these perspectives predict markedly different recovery shapes. Some anticipate widening gaps (cumulative advantage, cultural capital, tracking), others suggest possible narrowing in specific contexts (social capital, targeted policy), and still others predict complex or non-linear patterns (capability approach, intersectionality). Empirical testing of these hypotheses can illuminate not only whether educational outcomes recover, but also whether recovery is equitable and whether the pandemic's inequalities become embedded in long-term stratification patterns.

**Present Study**

Against this backdrop, the present study examines the impact of COVID-19-related school closures on socio-economic inequalities in secondary education outcomes in the Netherlands and their persistence in the years following the pandemic. While early research has illuminated learning losses in education in the short term, many of the studies highlighted the need to monitor long-term consequences on students' knowledge and skills, particularly whether pre-existing gaps in secondary schooling have widened or narrowed during the recovery period (Hanushek & Woessmann, 2020; Psacharopoulos et al., 2021). We address this question by analyzing comprehensive, registry-based data on national secondary school examinations from 2017 through 2023 – covering cohorts that graduated just before, during, and after the pandemic. This allows us to assess both the immediate fallout of school closures and the subsequent recovery trajectories up to three years post-crisis. We specifically focus on differential trends among socio-economic subgroups, leveraging indicators of parental education, household income, and other background factors available in Dutch administrative data. By integrating a subgroup analysis into a population-wide examination of learning outcomes, our study provides new evidence on whether COVID-19 exacerbated educational inequalities at the end of secondary education, and to what extent students from disadvantaged backgrounds have (or have not) caught up in the pandemic's aftermath. Ultimately, the findings aim to inform ongoing debates on educational equity in the pandemic recovery phase and to guide targeted interventions for those students who were left furthest behind.

## 2. METHODS:

*Data Source and Sample*

This study utilized administrative microdata obtained from Statistics Netherlands (Centraal Bureau voor de Statistiek; CBS), the national statistical office of the Netherlands. CBS collects and integrates a wide array of registry-based datasets on education, demographics, and socio-economic characteristics. Through a secure and pseudonymized data environment, the CBS provides high-quality longitudinal data for population-level research while ensuring strict adherence to privacy regulations under the General Data Protection Regulation (GDPR).

The analytical sample comprised 1,471,217 students who completed their final central examinations in one of the four major secondary education tracks in the Netherlands: pre-university education (VWO), senior general secondary education (HAVO), upper track of pre-vocational education (VMBO-TG), and the lower track of pre-vocational education (VMBO-BK). Students were included if they completed their final examinations in academic years spanning 2016–2017 to 2022–2023. For clarity, academic years are henceforth referred to by the calendar year in which final examinations were administered (e.g., “2021” refers to the 2020–2021 academic year). Table 1 lists characteristics of the sample and variables examined.

*Table 1. Characteristics of the analytical sample by educational track, including baseline cohorts (2017–2019) and pandemic cohorts (2020–2023), disaggregated by gender, migration background, parental education, household income, and urbanicity (N = 1,471,217)*

| | ***HAVO*** | ***VWO*** | ***VMBO_BK*** | ***VMBO_GT*** | ***Total*** |
|---|---|---|---|---|---|
| ***Group*** | | | | | |
| Control | 206,000 | 146,774 | 187,699 | 219,630 | 760,103 |
| 2019/2020 | 51,710 | 37,739 | 42,569 | 52,811 | 184,829 |
| 2020/2021 | 53,490 | 39,083 | 40,979 | 51,537 | 185,089 |
| 2021/2022 | 48,034 | 37,696 | 38,368 | 48,049 | 172,147 |
| 2022/2023 | 47,963 | 36,614 | 36,796 | 47,676 | 169,049 |
| ***Gender*** | | | | | |
| Female | 213,524 | 138,023 | 157,220 | 208,753 | 717,520 |
| Male | 193,673 | 159,883 | 189,191 | 210,950 | 753,697 |
| ***Migration Background*** | | | | | |
| Native | 326,089 | 242942 | 246,914 | 326,042 | 1,141,987 |
| 1st generation western | 4,827 | 4,081 | 5,890 | 60,888 | 75,686 |
| 2nd generation western | 21,451 | 19,459 | 14,114 | 19,801 | 74,825 |
| 1st generation non-western | 5,363 | 3,218 | 11,977 | 7,385 | 27,943 |
| 2nd generation non-western | 49,341 | 28,159 | 67,434 | 5,446 | 150,380 |
| ***Parental Education*** | | | | | |
| High | 84,129 | 116,299 | 12,771 | 39,999 | 253,198 |
| Middle | 212,291 | 125,336 | 156,257 | 205,576 | 699,460 |
| Low | 30,387 | 11,433 | 69,447 | 46,269 | 157,536 |
| Unknown | 80,390 | 44,838 | 107,936 | 127,859 | 361,023 |
| ***Urbanicity*** | | | | | |
| >=2500 houses/km2 | 85,809 | 94,515 | 68,031 | 78,705 | 327,060 |
| 1500 - 2500 houses/km2 | 129,365 | 92,602 | 103,343 | 128,405 | 453,715 |
| 1000 - 1500 houses/km2 | 68,433 | 43,282 | 55,752 | 70,635 | 238,102 |
| 500 - 1000 houses/km2 | 91,050 | 50,799 | 87,216 | 103,159 | 332,224 |
| <500 houses/km2 | 31,751 | 16,206 | 31,183 | 37,940 | 117,080 |
| ***Income*** | | | | | |
| Q1 (Highset) | 5,464 | 3,411 | 11,192 | 8,393 | 28,460 |
| Q2 (High) | 22,219 | 10,533 | 49,955 | 36,209 | 118,916 |
| Q3 (Middle) | 51,344 | 27,085 | 78,149 | 71,821 | 228,399 |
| Q4 (Low) | 112,063 | 64,434 | 111,036 | 135,656 | 423,189 |
| Q5 (Lowest) | 214,845 | 191,730 | 93,966 | 166,054 | 666,595 |
| Unknown | 1,262 | 713 | 2,113 | 1,570 | 5,658 |

*Outcome Measure*

The primary outcome variable was the student's total score on final examinations, aggregated across all subjects. This measure reflects standardized performance at the conclusion of secondary schooling and is used nationally for certification and higher education placement. There are two types of exams in secondary schools in the Netherlands, central exams and school exams. Each exam type has its own aggregate score, which was subsequently analyzed separately. No central exams took place in the academic year 2019-2020 due to COVID-19 related lockdown measures.

*Explanatory Variables*

The primary explanatory variable was students' exposure to COVID-19-related school closures. Exposure was operationalized through a set of indicator variables corresponding to the secondary school examination cohorts of 2020, 2021, 2022, and 2023, all of which experienced pandemic-related disruptions to teaching and assessment. Examination cohorts from 2017, 2018, and 2019 served as the pre-pandemic comparison group.

Several covariates were included to account for background characteristics. Gender was recorded as male or female. Migration background followed the classification of Statistics Netherlands (CBS), distinguishing between Dutch natives (no migration background), first-generation migrants (born abroad), and second-generation migrants (born in the Netherlands with at least one parent born abroad). Both first- and second-generation migrants were further disaggregated into Western and non-Western categories. In Dutch statistical practice, “Western” typically includes Europe (excluding Turkey), North America, Oceania, Indonesia, and Japan, while “non-Western” covers all other countries.

Parental education level was measured as the highest qualification attained by either parent. To facilitate international comparison, Dutch educational categories were mapped to three levels: low (primary education; lower tracks of Dutch secondary education, known as voorbereidend middelbaar beroepsonderwijs—vmbo—basis/kader; and entry-level vocational education, middelbaar beroepsonderwijs [MBO] level 1); medium (upper secondary general education, hoger algemeen voortgezet onderwijs—HAVO; pre-university education, voorbereidend wetenschappelijk onderwijs—VWO; intermediate and advanced vocational education, MBO levels 2–4; and bachelor’s degrees from either universities of applied sciences—hoger beroepsonderwijs [HBO]—or research universities—wetenschappelijk onderwijs [WO]); and high (master’s degrees or doctoral degrees from HBO or WO institutions). HBO refers to professionally oriented higher education, whereas WO refers to academic, research-oriented universities.

Parental income was defined as the average of the gross incomes of both parents, calculated over the relevant reference year, and divided into quintiles to capture relative economic position within the population.

Urbanicity was measured using a five-point ordinal scale representing the degree of urbanization of the student’s residential municipality, based on address density as defined by CBS. The scale ranges from 1 (very highly urbanized, ≥ 2,500 addresses per $km^2$) to 5 (very low urbanization, < 500 addresses per $km^2$).

*Analytical Strategy*

Analyses were stratified by educational track (VWO, HAVO, VMBO-TG, VMBO-BK) to account for differences in curricular content, examination standards, and student demographics. For each track, separate generalized linear models (GLM) were estimated to assess the association between COVID-19 exposure and examination performance.

All explanatory variables, except for gender, were categorical with more than two levels and were therefore included in the models using dummy coding. For each variable (e.g., parental education, migration background, parental income, and urbanicity), one category was used as the reference group, and binary indicators were created for the remaining categories. Gender was included as a binary indicator (male vs. female). All variables had limited missingness (<1%) except for parental educational attainment, where the data was more likely to be missing for students with migration background. In the case of parental education, unknown was coded as an additional level. For all other variables, individuals with missing values were excluded from the analysis.

A key analytical focus was on potential differential effects across socio-demographic groups. Accordingly, interaction terms between the exposure indicators and each SES-related variable (i.e. gender, migration background, parental education, income) were included in the models. This allowed the models to estimate changes in subgroup performance during pandemic-exposed years relative to the pre-pandemic baseline.

Using gender as an example, the general model specification was as follows:

$$Exam\ Score_i = \alpha + \beta_1 Exposure_i + \beta_2 * Gender_i + \beta_3 * Gender_i * Exposure_i + \beta_4 * Migration_i + \beta_5 * Parental\ education_i + \beta_6 * Parental\ Income_i + \beta_7 * Urbanicity_i + \varepsilon_i$$

where $\varepsilon_i$ denotes the individual error term. Analogous models were fitted for each SES dimension of interest and are reported in the annex. All analyses were conducted in a secure computing environment provided by CBS using Python. The code used to produce the analysis is available here {link redacted} and is indexed in the ODISSEI code library. Replication of the analysis environment is possible by contacting cbs at mircodata@cbs.nl and quoting project number 9898.

## 3. RESULTS

Table 2. Results of generalized linear models estimating exam performance by track and socio-demographic group, baseline and interaction effects

| | | Baseline | | | Interaction Effects | | | | | | | | |
|---|---|---|---|---|---|---|---|---|---|---|---|---|---|
| | | 2015-2019 | | | 2021 | | | 2022 | | | 2023 | | |
| | | **β** | **std err** | ***p*** | **β** | **std err** | ***p*** | **β** | **std err** | ***p*** | **β** | **std err** | ***p*** |
| | **VWO** | | | | | | | | | | | | |
| Overall Exposure | General Exposure | - | - | - | -0.07 | 0.01 | 0.00 | -0.14 | 0.01 | 0.00 | 0.02 | 0.01 | 0.00 |
| Gender (ref = man) | Woman | -0.16 | 0.01 | 0.00 | 0.04 | 0.01 | 0.00 | 0.03 | 0.01 | 0.02 | 0.03 | 0.01 | 0.01 |
| Migration Background (ref = Dutch native) | 1st Gen. Western | 0.09 | 0.02 | 0.00 | 0.01 | 0.05 | 0.90 | 0.06 | 0.05 | 0.25 | -0.01 | 0.05 | 0.86 |
| | 2nd Gen. Western | -0.03 | 0.01 | 0.00 | 0.00 | 0.02 | 0.93 | 0.00 | 0.02 | 0.98 | 0.04 | 0.02 | 0.07 |
| | 1st Gen. Non-western | -0.37 | 0.03 | 0.00 | -0.06 | 0.06 | 0.32 | 0.06 | 0.05 | 0.27 | 0.16 | 0.05 | 0.00 |
| | 2nd Gen. Non-western | -0.34 | 0.01 | 0.00 | -0.05 | 0.02 | 0.01 | -0.02 | 0.02 | 0.27 | -0.03 | 0.02 | 0.19 |
| Parental Education (ref = Middle) | High | 0.22 | 0.01 | 0.00 | -0.01 | 0.01 | 0.23 | 0.00 | 0.01 | 0.90 | 0.03 | 0.01 | 0.02 |
| | Low | -0.21 | 0.01 | 0.00 | -0.04 | 0.03 | 0.23 | -0.03 | 0.03 | 0.40 | -0.02 | 0.03 | 0.62 |
| Income (ref = middle quintile Q3) | Highest (Q1) | 0.00 | 0.03 | 0.90 | -0.11 | 0.06 | 0.07 | 0.06 | 0.06 | 0.32 | 0.11 | 0.06 | 0.07 |
| | High (Q2) | -0.06 | 0.02 | 0.00 | -0.04 | 0.04 | 0.33 | 0.00 | 0.04 | 0.92 | 0.00 | 0.04 | 0.98 |
| | Low (Q4) | 0.05 | 0.01 | 0.00 | 0.00 | 0.02 | 0.87 | 0.04 | 0.02 | 0.06 | 0.00 | 0.02 | 0.97 |
| | Lowest (Q5) | 0.08 | 0.01 | 0.00 | 0.00 | 0.02 | 0.84 | 0.03 | 0.02 | 0.15 | -0.01 | 0.02 | 0.67 |
| Urbanicity (ref = > 2,500 houses/km2) | 1,500 - 2,500 houses/km2 | -0.11 | 0.01 | 0.00 | 0.11 | 0.02 | 0.00 | 0.08 | 0.02 | 0.00 | 0.08 | 0.02 | 0.00 |
| | 1,000 - 1,500 houses/km2 | -0.17 | 0.01 | 0.00 | 0.17 | 0.02 | 0.00 | 0.19 | 0.02 | 0.00 | 0.16 | 0.02 | 0.00 |
| | 500 - 1,000 houses/km2 | -0.20 | 0.01 | 0.00 | 0.20 | 0.02 | 0.00 | 0.19 | 0.02 | 0.00 | 0.22 | 0.02 | 0.00 |
| | <500 houses/km2 | -0.23 | 0.01 | 0.00 | 0.20 | 0.03 | 0.00 | 0.23 | 0.03 | 0.00 | 0.21 | 0.03 | 0.00 |
| | **HAVO** | | | | | | | | | | | | |
| Overall Exposure | General Exposure | - | - | - | -0.08 | 0.01 | 0.00 | -0.14 | 0.01 | 0.00 | -0.01 | 0.01 | 0.18 |
| Gender (ref = man) | Woman | -0.22 | 0.00 | 0.00 | 0.10 | 0.01 | 0.00 | 0.07 | 0.01 | 0.00 | 0.00 | 0.01 | 0.89 |
| Migration Background (ref = Dutch native) | 1st Gen. Western | -0.04 | 0.02 | 0.11 | -0.18 | 0.05 | 0.00 | -0.06 | 0.05 | 0.22 | -0.08 | 0.04 | 0.08 |
| | 2nd Gen. Western | -0.05 | 0.01 | 0.00 | -0.09 | 0.02 | 0.00 | -0.02 | 0.02 | 0.32 | -0.03 | 0.02 | 0.24 |
| | 1st Gen. Non-western | -0.40 | 0.02 | 0.00 | -0.17 | 0.04 | 0.00 | -0.10 | 0.04 | 0.03 | -0.16 | 0.04 | 0.00 |
| | 2nd Gen. Non-western | -0.33 | 0.01 | 0.00 | -0.13 | 0.02 | 0.00 | -0.09 | 0.02 | 0.00 | -0.14 | 0.02 | 0.00 |
| Parental Education (ref = Middle) | High | 0.13 | 0.01 | 0.00 | 0.02 | 0.01 | 0.23 | 0.02 | 0.01 | 0.07 | 0.02 | 0.01 | 0.10 |
| | Low | -0.11 | 0.01 | 0.00 | -0.07 | 0.02 | 0.00 | -0.03 | 0.02 | 0.13 | -0.07 | 0.02 | 0.00 |
| Income (ref = middle quintile Q3) | Highest (Q1) | -0.04 | 0.02 | 0.01 | -0.03 | 0.05 | 0.50 | 0.02 | 0.05 | 0.64 | -0.08 | 0.05 | 0.09 |
| | High (Q2) | 0.04 | 0.02 | 0.05 | -0.05 | 0.03 | 0.07 | -0.02 | 0.03 | 0.51 | -0.05 | 0.03 | 0.07 |
| | Low (Q4) | -0.01 | 0.01 | 0.32 | 0.05 | 0.02 | 0.01 | 0.05 | 0.02 | 0.01 | 0.03 | 0.02 | 0.09 |
| | Lowest (Q5) | -0.01 | 0.01 | 0.41 | 0.06 | 0.02 | 0.00 | 0.04 | 0.02 | 0.01 | 0.05 | 0.02 | 0.00 |
| Urbanicity (ref = > 2,500 houses/km2) | 1,500 - 2,500 houses/km2 | -0.04 | 0.01 | 0.00 | 0.10 | 0.01 | 0.00 | 0.10 | 0.01 | 0.00 | 0.09 | 0.01 | 0.00 |
| | 1,000 - 1,500 houses/km2 | -0.07 | 0.01 | 0.00 | 0.15 | 0.02 | 0.00 | 0.13 | 0.02 | 0.00 | 0.14 | 0.02 | 0.00 |
| | 500 - 1,000 houses/km2 | -0.05 | 0.01 | 0.00 | 0.19 | 0.02 | 0.00 | 0.15 | 0.02 | 0.00 | 0.18 | 0.02 | 0.00 |
| | <500 houses/km2 | -0.07 | 0.01 | 0.00 | 0.21 | 0.02 | 0.00 | 0.19 | 0.02 | 0.00 | 0.18 | 0.02 | 0.00 |
| **VMBO-GT** | | | | | | | | | | | | | |
| Overall Exposure | General Exposure | - | - | - | -0.16 | 0.01 | 0.00 | -0.22 | 0.01 | 0.00 | 0.07 | 0.01 | 0.00 |
| Gender (ref = man) | Woman | -0.03 | 0.00 | 0.00 | -0.05 | 0.01 | 0.00 | -0.11 | 0.01 | 0.00 | -0.04 | 0.01 | 0.00 |
| Migration Background (ref = Dutch native) | 1st Gen. Western | 0.05 | 0.02 | 0.03 | -0.01 | 0.04 | 0.89 | 0.04 | 0.04 | 0.35 | -0.15 | 0.04 | 0.00 |
| | 2nd Gen. Western | -0.03 | 0.01 | 0.00 | -0.01 | 0.02 | 0.78 | 0.02 | 0.03 | 0.47 | -0.05 | 0.02 | 0.03 |
| | 1st Gen. Non-western | -0.30 | 0.02 | 0.00 | 0.08 | 0.04 | 0.03 | 0.07 | 0.04 | 0.05 | -0.16 | 0.04 | 0.00 |
| | 2nd Gen. Non-western | -0.28 | 0.01 | 0.00 | -0.05 | 0.01 | 0.00 | -0.01 | 0.02 | 0.36 | -0.13 | 0.02 | 0.00 |
| Parental Education (ref = Middle) | High | 0.17 | 0.01 | 0.00 | 0.03 | 0.02 | 0.07 | 0.01 | 0.02 | 0.38 | 0.10 | 0.04 | 0.03 |
| | Low | -0.10 | 0.01 | 0.00 | -0.04 | 0.02 | 0.01 | -0.04 | 0.02 | 0.01 | -0.04 | 0.04 | 0.28 |
| Income (ref = middle quintile Q3) | Highest (Q1) | -0.01 | 0.02 | 0.72 | -0.01 | 0.04 | 0.81 | -0.03 | 0.04 | 0.51 | -0.05 | 0.04 | 0.19 |
| | High (Q2) | 0.01 | 0.01 | 0.18 | -0.04 | 0.02 | 0.03 | 0.00 | 0.02 | 0.83 | -0.08 | 0.02 | 0.00 |
| | Low (Q4) | 0.02 | 0.01 | 0.02 | 0.01 | 0.02 | 0.60 | 0.01 | 0.02 | 0.47 | 0.03 | 0.02 | 0.09 |
| | Lowest (Q5) | 0.03 | 0.01 | 0.00 | 0.02 | 0.01 | 0.12 | 0.01 | 0.02 | 0.60 | 0.05 | 0.02 | 0.00 |
| Urbanicity (ref = > 2,500 houses/km2) | 1,500 - 2,500 houses/km2 | 0.01 | 0.01 | 0.12 | 0.03 | 0.02 | 0.05 | 0.02 | 0.02 | 0.20 | 0.05 | 0.02 | 0.00 |
| | 1,000 - 1,500 houses/km2 | 0.04 | 0.01 | 0.00 | 0.01 | 0.02 | 0.49 | -0.02 | 0.02 | 0.36 | 0.04 | 0.02 | 0.04 |
| | 500 - 1,000 houses/km2 | 0.07 | 0.01 | 0.00 | 0.05 | 0.02 | 0.00 | 0.02 | 0.02 | 0.12 | 0.08 | 0.02 | 0.00 |
| | <500 houses/km2 | 0.00 | 0.01 | 0.74 | 0.07 | 0.02 | 0.00 | 0.06 | 0.02 | 0.00 | 0.13 | 0.02 | 0.00 |
| **VMBO-BK** | | | | | | | | | | | | | |
| Overall Exposure | General Exposure | - | - | - | -0.20 | 0.01 | 0.00 | -0.23 | 0.01 | 0.00 | -0.06 | 0.01 | 0.00 |
| Gender (ref = man) | Woman | -0.17 | 0.01 | 0.00 | -0.09 | 0.01 | 0.00 | -0.07 | 0.01 | 0.00 | 0.03 | 0.01 | 0.01 |
| Migration Background (ref = Dutch native) | 1st Gen. Western | -0.09 | 0.02 | 0.00 | 0.00 | 0.04 | 0.95 | 0.00 | 0.04 | 0.93 | -0.03 | 0.04 | 0.44 |
| | 2nd Gen. Western | -0.06 | 0.01 | 0.00 | 0.03 | 0.03 | 0.30 | 0.04 | 0.03 | 0.19 | -0.07 | 0.03 | 0.03 |
| | 1st Gen. Non-western | -0.34 | 0.02 | 0.00 | -0.08 | 0.03 | 0.01 | -0.15 | 0.03 | 0.00 | -0.20 | 0.03 | 0.00 |
| | 2nd Gen. Non-western | -0.23 | 0.01 | 0.00 | -0.07 | 0.02 | 0.00 | -0.07 | 0.02 | 0.00 | -0.14 | 0.02 | 0.00 |
| Parental Education (ref = Middle) | High | 0.06 | 0.01 | 0.00 | 0.09 | 0.03 | 0.00 | 0.03 | 0.03 | 0.26 | -0.17 | 0.09 | 0.07 |
| | Low | -0.04 | 0.01 | 0.00 | -0.03 | 0.02 | 0.09 | -0.06 | 0.02 | 0.00 | -0.14 | 0.05 | 0.00 |
| Income (ref = middle quintile Q3) | Highest (Q1) | 0.01 | 0.01 | 0.39 | 0.00 | 0.03 | 0.94 | 0.00 | 0.04 | 0.96 | 0.01 | 0.04 | 0.87 |
| | High (Q2) | 0.02 | 0.01 | 0.04 | 0.03 | 0.02 | 0.10 | 0.01 | 0.02 | 0.73 | -0.03 | 0.02 | 0.10 |
| | Low (Q4) | -0.02 | 0.01 | 0.01 | 0.02 | 0.02 | 0.30 | 0.03 | 0.02 | 0.10 | 0.03 | 0.02 | 0.04 |
| | Lowest (Q5) | -0.02 | 0.01 | 0.00 | 0.02 | 0.02 | 0.15 | 0.03 | 0.02 | 0.06 | 0.05 | 0.02 | 0.00 |
| Urbanicity (ref = > 2,500 houses/km2) | 1,500 - 2,500 houses/km2 | 0.01 | 0.01 | 0.26 | 0.04 | 0.02 | 0.01 | 0.06 | 0.02 | 0.00 | 0.07 | 0.02 | 0.00 |
| | 1,000 - 1,500 houses/km2 | 0.02 | 0.01 | 0.01 | 0.08 | 0.02 | 0.00 | 0.02 | 0.02 | 0.30 | 0.06 | 0.02 | 0.00 |
| | 500 - 1,000 houses/km2 | -0.01 | 0.01 | 0.18 | 0.04 | 0.02 | 0.04 | 0.03 | 0.02 | 0.07 | 0.12 | 0.02 | 0.00 |
| | <500 houses/km2 | -0.01 | 0.01 | 0.06 | -0.01 | 0.02 | 0.80 | 0.01 | 0.02 | 0.69 | 0.10 | 0.03 | 0.00 |

This study examined the impact of COVID-19 school closures on central exam performance across four secondary education tracks in the Netherlands. Although the magnitude and recovery of learning losses varied by track, consistent patterns emerged when results are delineated by key socio-demographic variables. Table 2 summarizes the results of the GLM model for each track.

**Gender**

Across all tracks, girls scored lower than boys at baseline, though the size of the gender gap varied. During the pandemic, gender disparities reduced in the general education tracks (HAVO and VWO) and widened in the vocational tracks (VMBO-GT and VMBO-BK), particularly in 2021 and 2022. However, by 2023, the pandemic related gender gap narrowed in all tracks, returning to pre-pandemic levels.

**Migration Background**

Migration background had a stratified effect depending on the country of origin. While students from western countries of origins had no consistent disparities compared to students with a Dutch background, students with a non-Western migration background, especially first-generation students, were consistently and substantially disadvantaged in all tracks before the pandemic (e.g. VWO: -.37; HAVO: -.40; VMBO-GT: -.34). These baseline disparities intensified during the pandemic, particularly in HAVO, VMBO-GT and VMBO-BK, where first-generation non-Western migrant students experienced additional losses of 0.13 to 0.20 standard units by 2023. Second-generation non-Western migrant students also experienced similar compounding disadvantages, though the effect sizes were somewhat smaller. In contrast, first generation non-western migrant students in the VWO track showed some signs of resilience in the years following the pandemic, with significant reductions to the pre-pandemic performance gap by the year 2023, suggesting that migrant students in more academically oriented tracks may have had greater compensatory resources or institutional support.

**Parental Education**

Parental education level was a robust predictor of academic performance across all tracks. Students with highly educated parents consistently scored higher at baseline. During the pandemic, this advantage was preserved or even slightly widened in all four tracks. Conversely, students with lower parental education exhibited persistent or worsening performance gaps, particularly in HAVO and VMBO-BK, highlighting enduring structural inequities exacerbated by the pandemic.

**Household Income**

Household income effects were weaker and less consistent than those of parental education. In VWO track, students from lower income households tended to score better at baseline by 0.05-0.08 standard units, but there was no clear pattern of effect following school closures. In other tracks, the differences between quantiles at baseline or following exposure rarely exceeded 0.04 standard units.

**Urbanicity**

Urbanicity emerged as a surprising stratifier of pandemic related differences in school outcomes. In the VWO track, students from less urbanized areas (<2,500 houses/km$^2$) had worse scores at baseline, with an effect size that was proportional how rural their area was (e.g. β = -0.11 for second highest urbanicity and β = -0.23 for the lowest). However, following school closures, the baseline performance gap completely disappeared as the exposure interaction term was almost exactly the reverse of the difference at baseline, with a corresponding effect size that was proportional to the urbanicity rank. This effect was persistent in the three years following exposure. In the HAVO track, students from less urbanized areas also had worse scores at baseline compared to the highest urbanicity areas, though the effect size was comparable across all ranks. Nonetheless, the interaction terms were very similar to VWO track, showing better exam performance with lower urbanicity, with students from rural areas outperforming their urban peers in the three years following exposure. In the vocational tracks, effects of urbanicity were not consistent at baseline, yet rural regions still showed a relative advantage after exposure.

To provide an integrated overview of these subgroup patterns across tracks, Figure 1 plots predicted exam performance relative to the pre-pandemic baseline (2015–2019) for cohorts exposed to COVID-19 disruptions (2021–2023). Each column corresponds to one educational track (VWO, HAVO, VMBO-GT, VMBO-BK) and illustrates trajectories by gender, migration background, parental education, household income, and urbanicity in each row. The figure highlights both persistent inequalities (e.g., sustained disadvantages for students with lower parental education and for those with a non-Western migration background) and surprising reversals (e.g., rural students narrowing or exceeding urban peers post-pandemic)

Figure 1. Predicted exam performance relative to the pre-pandemic baseline (2015–2019) for cohorts exposed to COVID-19 disruptions (2021–2023). Columns correspond to educational tracks (VWO, HAVO, VMBO-GT, VMBO-BK); rows show trajectories by gender, migration background, parental education, household income, and urbanicity.

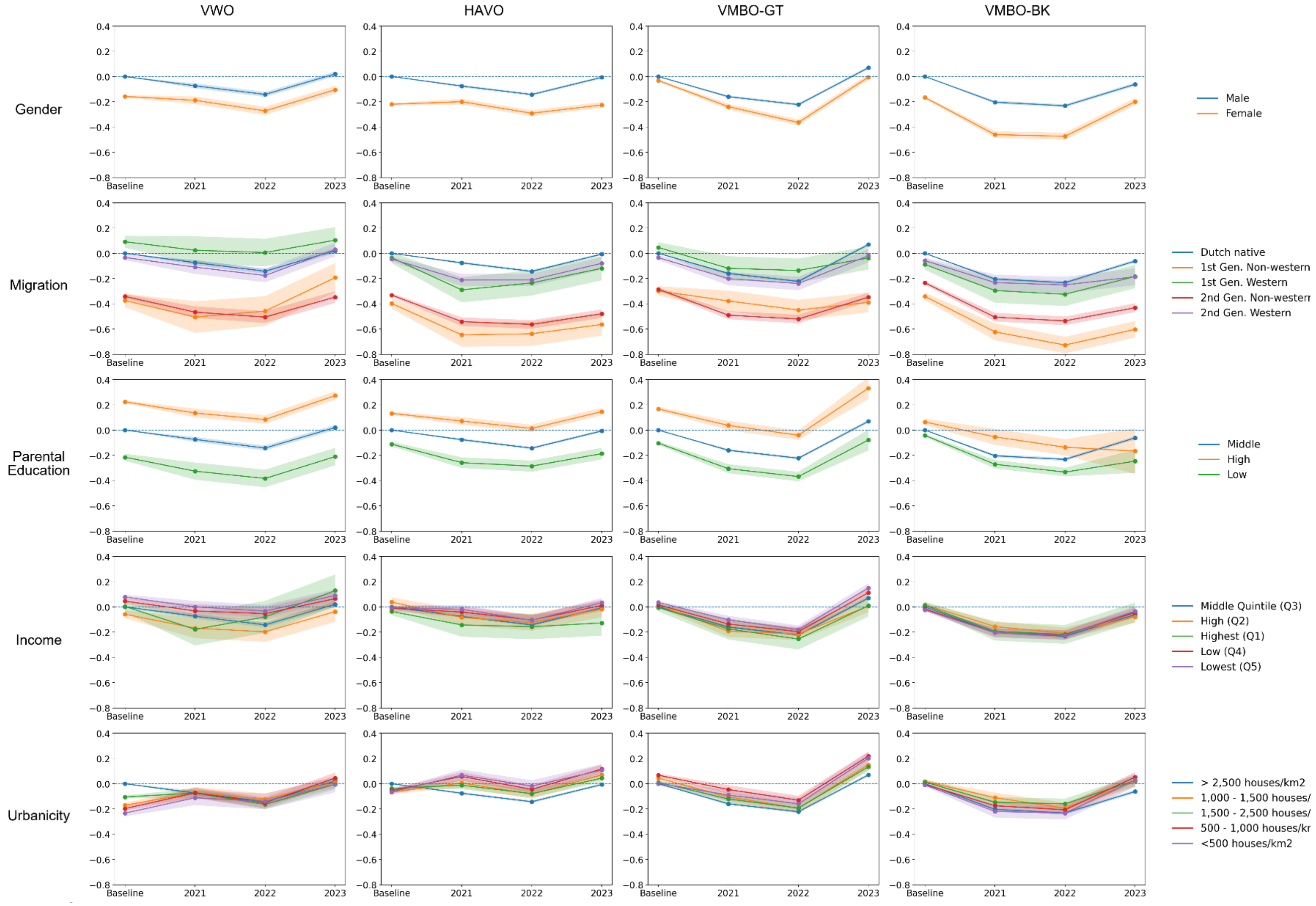

## 4. DISCUSSION:

This study provides robust evidence on the lasting and unequal effects of the COVID-19 pandemic on secondary school performance in the Netherlands. Using comprehensive administrative data covering over 1.4 million students across seven exam cohorts, we assessed the trajectory of central exam performance before, during, and after the pandemic. Our analysis focused on structural inequalities, revealing how pre-existing disparities—especially along lines of parental education, migration background, and urbanicity—shaped both the magnitude of learning loss and the speed of recovery.

**Persistent Inequalities and Uneven Recovery**

The results show that pandemic-related learning losses were not evenly distributed. While average exam performance declined in the years immediately following school closures, subgroup analyses indicate that these losses were significantly larger—and more persistent—for students from disadvantaged backgrounds. These findings resonate with prior international and Dutch studies showing disproportionate impacts of COVID-19 on children from less advantaged households (Betthäuser et al., 2023; Haelermans et al., 2022; Engzell et al., 2021). Our study adds to this evidence base by tracking cohorts up to 2023, revealing that some disparities not only remained but deepened several years after the initial disruption.

Students with a non-Western migration background—especially first-generation migrants—faced steep academic penalties during the pandemic. These students entered the crisis already behind, and the subsequent years saw widening gaps in most educational tracks.

The findings warrant investigation into systemic vulnerabilities, such as language barriers and reduced resources at home, that likely constrained their ability to benefit from remote learning or catch-up programs (Golden et al., 2023; Chen et al., 2022). The partial rebound seen among VWO students with a non-Western background by 2023 may point to the buffering role of academically selective tracks or targeted support in more elite school settings—an area warranting further qualitative exploration.

Parental education emerged as one of the more stable stratifiers across the entire study period. Students with lower-educated parents consistently scored lower, and the pandemic did not disrupt this pattern—if anything, it reinforced it. Unlike income, which showed inconsistent effects, parental education likely reflects cultural capital and home-based educational support, both of which became more critical during periods of remote or hybrid learning. These results are in line with previous work emphasizing the salience of non-monetary resources in educational resilience (Haelermans, Jacobs, et al., 2022).

**Surprising Moderators: Gender and Geography**

Two findings deserve special attention. First, gender effects were track-dependent: although girls generally scored lower at baseline, the gap fluctuated in complex ways across pandemic years and educational levels. In vocational tracks, girls were hit harder initially but demonstrated faster recovery by 2023. This raises important questions about gendered differences in learning environments, motivation, and school re-engagement in the aftermath of disruption.

Second, urbanicity showed a counterintuitive effect: students in rural and semi-rural areas performed worse before the pandemic but saw stronger recovery trajectories afterwards, effectively closing or even reversing the baseline urban advantage. This could reflect several mechanisms—including fewer school closures in rural areas, stronger community-school ties, or greater flexibility in local implementation of learning recovery programs. These hypotheses should be tested directly in future research that integrates school-level response data with local epidemiological and policy contexts.

The observed recovery patterns align with multiple theoretical perspectives on educational inequality and post-shock trajectories. Persistent or widening gaps, such as those for students with a non-Western migration background in HAVO and VMBO and for students with lower parental education in all tracks, are consistent with cumulative disadvantage and Matthew effects, where pre-existing disparities compound over time, slowing recovery for already disadvantaged groups. Cultural capital and inequality reproduction theories similarly predict that families with greater educational and cultural resources can better navigate post-pandemic recovery, preserving their relative advantage. The persistence or widening of these gaps also reflects relative risk aversion and effectively maintained inequality (EMI), whereby advantaged groups actively protect their status, while disadvantaged groups lack the means to compensate for pandemic-induced losses.

Track-dependent resilience, as seen in the narrowing of the gap for first-generation non-Western migrant students in VWO, aligns with tracking and compensatory advantage theories. Here, the high-status, academically selective nature of VWO may have offered stronger institutional support, more robust peer networks, and better-aligned recovery interventions. Institutional buffering further explains how elite tracks can shield disadvantaged students from the full effects of systemic shocks, providing structured opportunities to recover that are absent in lower-status tracks.

Gap closure to baseline, evident in the gender differences returning to pre-pandemic levels across all tracks by 2023, fits with education production models that predict convergence when normal learning conditions are restored. Once in-person schooling resumed and learning time equalized, gender disparities realigned with baseline patterns. This outcome may also reflect policy design effects if interventions, resources, or curriculum adjustments were applied evenly across genders, producing symmetrical recovery rates and preventing sustained divergence.

Gap reversal, most notably in the case of urbanicity—where rural students, initially behind, outperformed urban peers for three years post-pandemic—fits with social capital and capability approach perspectives. Stronger community-school ties, smaller class sizes, and greater continuity of schooling in rural areas could have facilitated stronger recovery, while urban students may have faced prolonged disruptions or less personal engagement. From a capability perspective, the reversal suggests that local conditions in rural settings may have expanded students’ real opportunities to learn post-pandemic, even if their absolute resource levels were lower.

Taken together, these patterns show that post-pandemic recovery was not uniform but instead reflected a combination of structural inequality reproduction and context-specific resilience. Some gaps proved resistant to change, consistent with long-term stratification processes, while others narrowed or even reversed where institutional or contextual factors provided buffers. The interplay between family resources, school structures, and local environments appears to have decisively shaped the shape and speed of educational recovery.

**Contributions, Limitations, and Future Directions**

This study offers three major contributions. First, it extends the time horizon of pandemic impact research into the post-crisis years, demonstrating that inequalities not only persisted but evolved. Second, it uses high-quality, population-level administrative data, overcoming many of the sampling and reporting limitations of survey-based studies. Third, it disaggregates trends across both academic and vocational tracks, revealing that pandemic effects were stratified not just by SES but also by educational pathway.

Nevertheless, some limitations merit discussion. Our results could not definitively identify learning loss on the population level following school closures, as our outcome variable (centralized exams) in the Netherlands are calibrated and normalized every year based on the overall performance of the students. Furthermore, our models cannot fully account for unobserved school-level heterogeneity, such as differences in the implementation of remote learning, teacher absence, or school leadership during the pandemic. While it was possible to use multilevel modelling to control for clustering of effects within certain schools, doing so would have masked effects of migration background and urbanicity, as both of these variables show clustering within schools. Furthermore, the analysis focuses on academic performance on centralized (final) exams, and does not capture other dimensions of student development, such as well-being, socio-emotional skills, or dropout risk. Finally, while our models test interactions with key background variables, they remain observational and cannot definitively establish causal mechanisms.

Future work should explore individual-level learning trajectories using longitudinal methods—for example, comparing standardized test scores from early primary school (e.g., CITO) to final secondary outcomes. This would allow for better estimation of intra-individual change and more precise identification of which students fell behind and why. Additionally, qualitative studies with students, parents, and educators could deepen our understanding of the lived experience behind the observed patterns—particularly in subgroups that defied expectations, such as rural students or low-income students in VWO. Ultimately, understanding the long term consequences  of the pandemic on social inequalities will take a long time as we continue to track and observe those groups affected.

**Implications for Equity-Oriented Policy**

The implications of these findings are both urgent and long-term. If pandemic-era learning losses are left unaddressed, they risk translating into structural disadvantages in higher education, employment, and social mobility, especially for already marginalized students (Maldonado et al., 2024; Psacharopoulos et al., 2021). Educational recovery programs must go beyond “average” remediation efforts and explicitly target those subgroups who have been left furthest behind. In practice, this may mean differentiated instruction, school-level investment, culturally responsive pedagogy, and tailored mentoring and support in vocational tracks. Moreover, the apparent resilience in some rural areas suggests that smaller, community-integrated schools may offer valuable lessons for system-wide recovery strategies.

In sum, the COVID-19 pandemic was not only a public health crisis but also a stress test for educational equity. While Dutch students have begun to recover on average, the gains have been far from evenly shared. Ensuring that recovery is not just statistical but substantive—and that no student cohort carries forward the scars of this disruption—must remain a national priority.

# 1 Annex 1 –

| | | Baseline | | | Interaction Effects | | | | | | | | | | | |
|---|---|---|---|---|---|---|---|---|---|---|---|---|---|---|---|---|
| | | 2015-2019 | | | 2020 | | | 2021 | | | 2022 | | | 2023 | | |
| | | β | std err | *p* | β | std err | *p* | β | std err | *p* | β | std err | *p* | β | std err | *p* |
| VWO | | | | | | | | | | | | | | | | |
| Overall Exposure | General Exposure | - | - | - | 0.22 | 0.01 | 0.00 | -0.02 | 0.01 | 0.00 | -0.10 | 0.01 | 0.00 | -0.09 | 0.01 | 0.00 |
| Gender (ref = man) | Woman | 0.23 | 0.01 | 0.00 | 0.09 | 0.01 | 0.00 | 0.04 | 0.01 | 0.00 | -0.01 | 0.01 | 0.37 | -0.03 | 0.01 | 0.00 |
| Migration Background (ref = Dutch native) | 1st Gen. Western | 0.16 | 0.02 | 0.00 | -0.10 | 0.05 | 0.07 | -0.03 | 0.05 | 0.59 | 0.04 | 0.05 | 0.47 | -0.01 | 0.05 | 0.76 |
| | 2nd Gen. Western | 0.00 | 0.01 | 0.75 | 0.01 | 0.02 | 0.73 | -0.02 | 0.02 | 0.45 | 0.03 | 0.02 | 0.25 | 0.04 | 0.02 | 0.07 |
| | 1st Gen. Non-western | -0.15 | 0.03 | 0.00 | 0.22 | 0.06 | 0.00 | -0.01 | 0.06 | 0.86 | 0.11 | 0.05 | 0.04 | 0.20 | 0.05 | 0.00 |
| | 2nd Gen. Non-western | -0.22 | 0.01 | 0.00 | 0.04 | 0.02 | 0.03 | -0.02 | 0.02 | 0.28 | 0.00 | 0.02 | 0.91 | -0.03 | 0.02 | 0.16 |
| Parental Education (ref = Middle) | High | 0.18 | 0.01 | 0.00 | -0.06 | 0.01 | 0.00 | 0.00 | 0.01 | 0.77 | 0.01 | 0.01 | 0.50 | 0.04 | 0.01 | 0.00 |
| | Low | -0.14 | 0.01 | 0.00 | 0.01 | 0.03 | 0.64 | -0.02 | 0.03 | 0.53 | -0.04 | 0.03 | 0.15 | -0.01 | 0.03 | 0.82 |
| Income (ref = middle quintile Q3) | Highest (Q1) | -0.01 | 0.02 | 0.80 | 0.09 | 0.06 | 0.12 | -0.05 | 0.06 | 0.44 | 0.09 | 0.06 | 0.15 | 0.08 | 0.06 | 0.17 |
| | High (Q2) | -0.05 | 0.02 | 0.00 | 0.06 | 0.04 | 0.13 | 0.01 | 0.04 | 0.77 | 0.02 | 0.04 | 0.66 | -0.02 | 0.04 | 0.68 |
| | Low (Q4) | 0.05 | 0.01 | 0.00 | 0.01 | 0.02 | 0.72 | -0.01 | 0.02 | 0.78 | 0.02 | 0.02 | 0.52 | -0.01 | 0.02 | 0.61 |
| | Lowest ((Q5) | 0.08 | 0.01 | 0.00 | -0.03 | 0.02 | 0.20 | 0.01 | 0.02 | 0.80 | 0.01 | 0.02 | 0.61 | -0.02 | 0.02 | 0.42 |
| Urbanicity (ref = > 2,500 houses/km2) | 1,500 - 2,500 houses/km2 | -0.10 | 0.01 | 0.00 | 0.07 | 0.02 | 0.00 | 0.09 | 0.02 | 0.00 | 0.06 | 0.02 | 0.00 | 0.08 | 0.02 | 0.00 |
| | 1,000 - 1,500 houses/km2 | -0.17 | 0.01 | 0.00 | 0.17 | 0.02 | 0.00 | 0.14 | 0.02 | 0.00 | 0.16 | 0.02 | 0.00 | 0.12 | 0.02 | 0.00 |
| | 500 - 1,000 houses/km2 | -0.19 | 0.01 | 0.00 | 0.14 | 0.02 | 0.00 | 0.15 | 0.02 | 0.00 | 0.13 | 0.02 | 0.00 | 0.18 | 0.02 | 0.00 |
| | <500 houses/km2 | -0.18 | 0.01 | 0.00 | 0.16 | 0.03 | 0.00 | 0.14 | 0.03 | 0.00 | 0.18 | 0.03 | 0.00 | 0.20 | 0.03 | 0.00 |
| HAVO | | | | | | | | | | | | | | | | |
| Overall Exposure | General Exposure | | | | 0.31 | 0.01 | 0.00 | -0.09 | 0.01 | 0.00 | -0.17 | 0.01 | 0.00 | -0.11 | 0.01 | 0.00 |
| Gender (ref = man) | Woman | 0.26 | 0.00 | 0.00 | 0.13 | 0.01 | 0.00 | 0.06 | 0.01 | 0.00 | 0.00 | 0.01 | 0.76 | -0.02 | 0.01 | 0.12 |
| Migration Background (ref = Dutch native) | 1st Gen. Western | 0.01 | 0.02 | 0.57 | -0.05 | 0.05 | 0.30 | -0.15 | 0.05 | 0.00 | -0.02 | 0.05 | 0.71 | -0.07 | 0.04 | 0.10 |
| | 2nd Gen. Western | -0.04 | 0.01 | 0.00 | -0.01 | 0.02 | 0.75 | -0.06 | 0.02 | 0.01 | 0.00 | 0.02 | 0.95 | 0.02 | 0.02 | 0.34 |
| | 1st Gen. Non-western | -0.20 | 0.02 | 0.00 | 0.15 | 0.05 | 0.00 | -0.04 | 0.04 | 0.30 | 0.04 | 0.04 | 0.35 | -0.04 | 0.04 | 0.33 |
| | 2nd Gen. Non-western | -0.21 | 0.01 | 0.00 | 0.07 | 0.02 | 0.00 | -0.09 | 0.02 | 0.00 | -0.05 | 0.02 | 0.00 | -0.11 | 0.02 | 0.00 |
| Parental Education (ref = Middle) | High | 0.10 | 0.01 | 0.00 | -0.04 | 0.01 | 0.00 | 0.03 | 0.01 | 0.04 | 0.04 | 0.01 | 0.00 | 0.01 | 0.01 | 0.48 |
| | Low | -0.07 | 0.01 | 0.00 | 0.03 | 0.02 | 0.14 | -0.05 | 0.02 | 0.00 | -0.04 | 0.02 | 0.07 | -0.07 | 0.02 | 0.00 |
| Income (ref = middle quintile Q3) | Highest (Q1) | 0.02 | 0.02 | 0.35 | -0.02 | 0.05 | 0.60 | -0.02 | 0.05 | 0.61 | -0.03 | 0.05 | 0.50 | -0.05 | 0.05 | 0.30 |
| | High (Q2) | -0.02 | 0.01 | 0.09 | 0.04 | 0.03 | 0.09 | -0.05 | 0.03 | 0.07 | 0.00 | 0.03 | 0.94 | -0.02 | 0.03 | 0.54 |
| | Low (Q4) | 0.02 | 0.01 | 0.01 | 0.00 | 0.02 | 0.82 | 0.02 | 0.02 | 0.17 | 0.03 | 0.02 | 0.16 | 0.02 | 0.02 | 0.34 |
| | Lowest ((Q5) | 0.02 | 0.01 | 0.00 | -0.02 | 0.02 | 0.21 | 0.03 | 0.02 | 0.03 | 0.01 | 0.02 | 0.65 | 0.00 | 0.02 | 0.78 |
| Urbanicity (ref = > 2,500 houses/km2) | 1,500 - 2,500 houses/km2 | 0.00 | 0.01 | 0.79 | 0.05 | 0.01 | 0.00 | 0.09 | 0.01 | 0.00 | 0.08 | 0.01 | 0.00 | 0.07 | 0.01 | 0.00 |
| | 1,000 - 1,500 houses/km2 | -0.02 | 0.01 | 0.01 | 0.08 | 0.02 | 0.00 | 0.11 | 0.02 | 0.00 | 0.05 | 0.02 | 0.00 | 0.07 | 0.02 | 0.00 |
| | 500 - 1,000 houses/km2 | 0.03 | 0.01 | 0.00 | 0.05 | 0.02 | 0.00 | 0.11 | 0.02 | 0.00 | 0.11 | 0.02 | 0.00 | 0.12 | 0.02 | 0.00 |
| | <500 houses/km2 | 0.03 | 0.01 | 0.00 | 0.03 | 0.02 | 0.13 | 0.15 | 0.02 | 0.00 | 0.10 | 0.02 | 0.00 | 0.13 | 0.02 | 0.00 |
| VMBO-GT | | | | | | | | | | | | | | | | |
| Overall Exposure | General Exposure | | | | 0.00 | 0.01 | 0.63 | -0.18 | 0.01 | 0.00 | -0.25 | 0.01 | 0.00 | -0.18 | 0.01 | 0.00 |
| Gender (ref = man) | Woman | 0.29 | 0.00 | 0.00 | 0.00 | 0.01 | 0.89 | -0.01 | 0.01 | 0.16 | -0.02 | 0.01 | 0.05 | -0.04 | 0.01 | 0.00 |
| Migration Background (ref = Dutch native) | 1st Gen. Western | 0.07 | 0.02 | 0.00 | -0.10 | 0.04 | 0.03 | 0.00 | 0.04 | 0.92 | -0.04 | 0.04 | 0.35 | -0.13 | 0.04 | 0.00 |
| | 2nd Gen. Western | -0.04 | 0.01 | 0.00 | -0.01 | 0.02 | 0.61 | -0.03 | 0.02 | 0.29 | 0.02 | 0.02 | 0.42 | -0.01 | 0.02 | 0.63 |
| | 1st Gen. Non-western | -0.10 | 0.02 | 0.00 | 0.11 | 0.04 | 0.01 | 0.10 | 0.04 | 0.00 | 0.07 | 0.04 | 0.04 | -0.07 | 0.03 | 0.03 |
| | 2nd Gen. Non-western | -0.19 | 0.01 | 0.00 | -0.03 | 0.01 | 0.01 | -0.03 | 0.01 | 0.03 | -0.04 | 0.01 | 0.00 | -0.11 | 0.01 | 0.00 |
| Parental Education (ref = Middle) | High | 0.13 | 0.01 | 0.00 | -0.03 | 0.02 | 0.03 | 0.04 | 0.02 | 0.01 | 0.02 | 0.02 | 0.31 | 0.09 | 0.04 | 0.03 |
| | Low | -0.07 | 0.01 | 0.00 | -0.03 | 0.02 | 0.09 | -0.05 | 0.02 | 0.00 | -0.07 | 0.02 | 0.00 | -0.02 | 0.04 | 0.70 |
| Income (ref = middle quintile Q3) | Highest (Q1) | -0.03 | 0.02 | 0.07 | -0.06 | 0.04 | 0.13 | -0.02 | 0.04 | 0.67 | -0.02 | 0.04 | 0.62 | -0.08 | 0.04 | 0.05 |
| | High (Q2) | -0.02 | 0.01 | 0.08 | -0.02 | 0.02 | 0.31 | -0.04 | 0.02 | 0.06 | -0.02 | 0.02 | 0.40 | -0.04 | 0.02 | 0.11 |
| | Low (Q4) | 0.04 | 0.01 | 0.00 | 0.01 | 0.01 | 0.64 | 0.01 | 0.02 | 0.46 | 0.02 | 0.02 | 0.26 | 0.02 | 0.02 | 0.32 |
| | Lowest ((Q5) | 0.05 | 0.01 | 0.00 | -0.01 | 0.01 | 0.55 | 0.02 | 0.01 | 0.13 | 0.02 | 0.02 | 0.26 | 0.02 | 0.02 | 0.15 |
| Urbanicity (ref = > 2,500 houses/km2) | 1,500 - 2,500 houses/km2 | 0.04 | 0.01 | 0.00 | 0.00 | 0.01 | 0.78 | -0.03 | 0.01 | 0.02 | -0.02 | 0.02 | 0.31 | -0.05 | 0.02 | 0.00 |
| | 1,000 - 1,500 houses/km2 | 0.04 | 0.01 | 0.00 | -0.02 | 0.02 | 0.24 | -0.05 | 0.02 | 0.00 | 0.00 | 0.02 | 0.80 | -0.02 | 0.02 | 0.35 |
| | 500 - 1,000 houses/km2 | 0.07 | 0.01 | 0.00 | -0.01 | 0.02 | 0.40 | -0.06 | 0.02 | 0.00 | -0.01 | 0.02 | 0.34 | 0.01 | 0.02 | 0.46 |
| | <500 houses/km2 | 0.05 | 0.01 | 0.00 | 0.05 | 0.02 | 0.01 | -0.03 | 0.02 | 0.17 | 0.04 | 0.02 | 0.05 | 0.02 | 0.02 | 0.39 |
| VMBO-BK | | | | | | | | | | | | | | | | |
| Overall Exposure | General Exposure | | | | 0.22 | 0.01 | 0.00 | -0.14 | 0.01 | 0.00 | -0.21 | 0.01 | 0.00 | -0.11 | 0.01 | 0.00 |
| Gender (ref = man) | Woman | 0.27 | 0.01 | 0.00 | 0.12 | 0.01 | 0.00 | -0.03 | 0.01 | 0.01 | -0.01 | 0.01 | 0.45 | -0.03 | 0.01 | 0.02 |
| Migration Background (ref = Dutch native) | 1st Gen. Western | -0.05 | 0.02 | 0.01 | 0.00 | 0.04 | 0.97 | -0.01 | 0.04 | 0.77 | 0.06 | 0.04 | 0.14 | 0.04 | 0.04 | 0.28 |
| | 2nd Gen. Western | -0.06 | 0.01 | 0.00 | 0.00 | 0.03 | 0.96 | 0.03 | 0.03 | 0.32 | 0.08 | 0.03 | 0.01 | -0.03 | 0.03 | 0.34 |
| | 1st Gen. Non-western | -0.12 | 0.02 | 0.00 | -0.02 | 0.03 | 0.54 | 0.07 | 0.03 | 0.03 | 0.05 | 0.03 | 0.10 | -0.06 | 0.03 | 0.02 |
| | 2nd Gen. Non-western | -0.22 | 0.01 | 0.00 | 0.00 | 0.01 | 0.93 | 0.03 | 0.01 | 0.02 | 0.04 | 0.02 | 0.02 | -0.08 | 0.02 | 0.00 |
| Parental Education (ref = Middle) | High | 0.13 | 0.01 | 0.00 | -0.04 | 0.03 | 0.16 | 0.05 | 0.03 | 0.07 | 0.05 | 0.03 | 0.09 | -0.23 | 0.09 | 0.01 |
| | Low | -0.09 | 0.01 | 0.00 | 0.01 | 0.01 | 0.64 | -0.02 | 0.01 | 0.24 | -0.02 | 0.02 | 0.12 | -0.03 | 0.04 | 0.47 |
| Income (ref = middle quintile Q3) | Highest (Q1) | -0.08 | 0.01 | 0.00 | 0.02 | 0.03 | 0.52 | 0.02 | 0.03 | 0.58 | 0.05 | 0.03 | 0.15 | 0.01 | 0.04 | 0.73 |
| | High (Q2) | -0.05 | 0.01 | 0.00 | 0.01 | 0.02 | 0.65 | 0.04 | 0.02 | 0.03 | 0.06 | 0.02 | 0.00 | -0.01 | 0.02 | 0.49 |
| | Low (Q4) | 0.05 | 0.01 | 0.00 | 0.03 | 0.02 | 0.06 | -0.01 | 0.02 | 0.50 | 0.01 | 0.02 | 0.59 | 0.00 | 0.02 | 0.92 |
| | Lowest ((Q5) | 0.05 | 0.01 | 0.00 | 0.03 | 0.02 | 0.08 | 0.01 | 0.02 | 0.60 | -0.01 | 0.02 | 0.78 | 0.00 | 0.02 | 0.98 |
| Urbanicity (ref = > 2,500 houses/km2) | 1,500 - 2,500 houses/km2 | 0.05 | 0.01 | 0.00 | -0.05 | 0.02 | 0.00 | -0.04 | 0.02 | 0.02 | -0.02 | 0.02 | 0.22 | -0.02 | 0.02 | 0.25 |
| | 1,000 - 1,500 houses/km2 | 0.08 | 0.01 | 0.00 | -0.05 | 0.02 | 0.01 | -0.06 | 0.02 | 0.00 | -0.04 | 0.02 | 0.02 | 0.02 | 0.02 | 0.41 |
| | 500 - 1,000 houses/km2 | 0.10 | 0.01 | 0.00 | -0.09 | 0.02 | 0.00 | -0.08 | 0.02 | 0.00 | -0.05 | 0.02 | 0.00 | 0.04 | 0.02 | 0.01 |
| | <500 houses/km2 | 0.09 | 0.01 | 0.00 | -0.07 | 0.02 | 0.00 | -0.08 | 0.02 | 0.00 | -0.08 | 0.02 | 0.00 | -0.01 | 0.02 | 0.66 |

1